\documentclass[]{aa} 
\usepackage{graphicx}
\usepackage{txfonts}
\usepackage{natbib}
\usepackage{ulem}
\usepackage{url}
\usepackage[usenames]{color}
\bibpunct{(}{)}{;}{a}{}{,} 
\newcount\longrefs
\def\aap{\ifnum\longrefs=1 {Astron.\ Astrophys.}\else 
                           {A\hbox{\rm \&}A}\fi}
\def\aapr{\ifnum\longrefs=1 {Astron.\ Astrophys.\ Rev.}\else 
                            {A\hbox{\rm \&}AR}\fi}
\def\aaps{\ifnum\longrefs=1 {Astron.\ Astrophys.\ Suppl.}\else 
                            {A\hbox{\rm \&}A Suppl.}\fi}
\def\aj{\ifnum\longrefs=1 {Astron.\ J.}\else 
                          {AJ}\fi} 
\def\ao{\ifnum\longrefs=1 {Applied Optics}\else 
                           {Appl.\ Opt.}\fi} 
\def\aspcs{\ifnum\longrefs=1 {Astron.\ Soc.\ Pacific Conf. Series}\else 
                           {ASP Conf.\ Ser.}\fi} 
\def\apj{\ifnum\longrefs=1 {Astrophys.\ J.}\else 
                           {ApJ}\fi} 
\def\apjl{\ifnum\longrefs=1 {Astrophys.\ J. Lett.}\else 
                            {ApJ}\fi} 
\def\aplett{\ifnum\longrefs=1 {Astrophys.\ J. Lett.}\else 
                            {ApJ}\fi} 
\def\apjs{\ifnum\longrefs=1 {Astrophys.\ J. Suppl.}\else 
                            {ApJS}\fi}
\def\apss{\ifnum\longrefs=1 {Astrophys.\ and Space Science}\else 
                            {Astrophys.\ Space Sci.}\fi}
\def\araa{\ifnum\longrefs=1 {Ann.\ Rev.\ Astron.\ Astrophys.}\else 
                            {ARA\hbox{\rm \&}A}\fi}
\def\azh{\ifnum\longrefs=1 {Astronomicheskii Zhurnal}\else 
                            {Astron.\ Zhur.}\fi}
\def\baas{\ifnum\longrefs=1 {Bull.\ Am.\ Astron.\ Soc.}\else 
                            {BAAS}\fi}
\def\bain{\ifnum\longrefs=1 {Bull.\ Astronom.\ Institutes Netherlands}\else
                            {Bull.\ Astr.\ Inst.\ Neth.}\fi}
\def\gca{\ifnum\longrefs=1 {Geochim.\ Cosmochim.\ Acta}\else 
                           {Geochim.\ Cosmochim.\ Acta}\fi}
\def\grl{\ifnum\longrefs=1 {Geophys.\ Res.\ Lett.}\else 
                           {Geoph.\ Res.\ Lett.}\fi}
\def\iaucirc{\ifnum\longrefs=1 {IAU Circulars}\else 
                          {IAU Circ.}\fi}
\def\ip{\ifnum\longrefs=1 {in press}\else 
                          {in press}\fi}
\def\jgr{\ifnum\longrefs=1 {J.\ Geophys.\ Res.}\else 
                           {J.\ Geophys.\ Res.}\fi}  
\def\jrasc{\ifnum\longrefs=1 {J.\ Royal Astron.\ Soc.\ Canada}\else 
                           {JRAS Can.}\fi}  
\def\mnras{\ifnum\longrefs=1 {Mon.\ Not.\ Roy.\ Astron.\ Soc.}\else 
                             {MNRAS}\fi} 
\def\nat{\ifnum\longrefs=1 {Nature}\else 
                           {Nat}\fi}
\def\pasj{\ifnum\longrefs=1 {Pub.\ Astron.\ Soc.\ Japan}\else 
                            {PASJ}\fi} 
\def\pasp{\ifnum\longrefs=1 {Pub.\ Astron.\ Soc.\ Pacific}\else 
                            {PASP}\fi} 
\def\physscr{\ifnum\longrefs=1 {Physica Scripta}\else 
                            {Phys.\ Scrip.}\fi} 
\def\planss{\ifnum\longrefs=1 {Planetary \& Space Science}\else 
                            {Plan. \& Space Sci.}\fi} 
\def\procspie{\ifnum\longrefs=1 {Proc.\ SPIE}\else 
                            {Proc.\ SPIE}\fi} 
\def\qjras{\ifnum\longrefs=1 {Quarterly J.\ Royal Astron.\ Soc.}\else 
                            {QJRAS}\fi} 
\def\sa{\ifnum\longrefs=1 {Soviet Astron..}\else 
                               {Sov.\ Astron.}\fi}
\def\skytel{\ifnum\longrefs=1 {Sky \& Telescope}\else 
                            {Sky \& Tel.}\fi} 
\def\solphys{\ifnum\longrefs=1 {Solar Phys.}\else 
                               {Sol.\ Phys.}\fi}
\def\ssr{\ifnum\longrefs=1 {Space Science Rev.}\else 
                               {Space\ Sci.\ Rev.}\fi}

\def\nl{,\ } 

\def\ARC{NASA Ames Research Center\nl Mof{}fett Field\nl CA~94035\nl USA}

\def\LPARL{Lockheed Martin ATC\nl Solar \& Astrophysics Lab\nl 
         Org.\ H1--12, Bldg.\ 252\nl 3251 Hanover Street\nl 
         Palo Alto, CA~94304--1187\nl USA}

\def\Oslo{Institute of Theoretical Astrophysics, University of Oslo\nl 
         P.O. Box 1029, Blindern\nl N--0315 Oslo\nl Norway}

\def\UUt{Utrecht University\nl Postbus 80\,000\nl
         NL--3508 TA Utrecht\nl The Netherlands}
\def\SPO{NSO/Sacramento Peak\nl P.O. Box 62\nl 
         Sunspot, NM 88349--0062\nl USA}

\def\dutch{\def\refname{Referenties}\def\abstractname{Samenvatting}%
  \def\bibname{Bibliografie}\def\chaptername{Hoofdstuk}%
  \def\appendixname{Bijlage}\def\contentsname{Inhoudsopgave}%
  \def\listfigurename{Lijst van figuren}%
  \def\listtablename{Lijst van tabellen}%
  \def\indexname{Index}\def\figurename{Figuur}\def\tablename{Tabel}%
  \def\partname{Deel}\def\enclname{Bijlage(n)}\def\ccname{Ter attentie van}%
  \def\headtoname{Aan}\def\headpagename{Pagina}%
  \def\today{\number\day\space\ifcase\month\or januari\or februari\or%
     maart\or%
     april\or mei\or juni\or juli\or augustus\or september\or oktober\or%
     november\or december\fi \space\number\year}%
  \typeout{
              >>>>> use hlatex209 for Dutch hyphenation <<<<< 
         }}

\DeclareFontFamily{OT1}{mvs}{}
\DeclareFontShape{OT1}{mvs}{m}{n}{<-> fmvr8x}{}

\newcounter{onefig} \newcounter{fignumber}
\newcount\nocaptions \newcount\nofigures \newcount\figwidth
\newcount\viewgraphs \newcount\printlabel
\long\def\skipfigure #1\viewout{}   
  \def\paper{}  \def\figlabel{} 
\long\def\nextfig#1{\setcounter{figure}{\value{fignumber}}
  \addtocounter{fignumber}{1}
  \ifnum \viewgraphs=1 \pagestyle{empty} \fi 
  \ifnum\value{onefig}=0 #1 \fi                 
  \ifnum\value{onefig}=\value{fignumber} #1 \fi}
\def\figwidths#1#2{\ifnum \nocaptions=1 #2mm \else #1mm \fi}  
\def\picplace{\framebox[80mm]{\rule{0cm}{1cm}}}
\def\paper#1{}  
\long\def\plotfig#1#2{\ifnum \nofigures=1 \picplace \else #2 \fi}
\long\def\captiontext#1{\ifnum \nofigures=1 \raggedright \fi 
   \ifnum \nocaptions=1 \paper
     \ifnum \viewgraphs=0 
       \newline  \mbox{}\hrulefill\mbox{} \newline 
       \ifnum \printlabel=1 \{{\em \figlabel}\}\newline \fi
     \fi 
   \else \ifnum \printlabel=1 \{{\em \figlabel}\}\newline \fi
     #1 \fi}

\newcount\panelwidth \newcount\panelheight 
\newcount\bxmin \newcount\bymin \newcount\bxmax \newcount\bymax
\newcount\tbxmin \newcount\tbymin
\newcount\tpanelwidth \newcount\tpanelheight \newcount\tpdif
\def\panelsize #1,#2;{\panelwidth=#1 \panelheight=#2}  
\def\setbb #1,#2;#3,#4;#5,#6;{
  \tbxmin=#1 \tbymin=#2    
  \bxmin=#3 \bymin=#4      
  \bxmax=#5 \bymax=#6}     
\def\barepanel #1{%
  \ifnum\panelheight=0 
    \tpdif=\bymax \advance\tpdif by -\bymin
    \multiply \tpdif by \panelwidth
    \tpanelheight=\tpdif
    \tpdif=\bxmax \advance\tpdif by -\bxmin
    \divide \tpanelheight by \tpdif
  \else \tpanelheight=\panelheight \fi
  \ifnum\panelwidth=0 
    \tpdif=\bxmax \advance\tpdif by -\bxmin
    \multiply \tpdif by \panelheight
    \tpanelwidth=\tpdif
    \tpdif=\bymax \advance\tpdif by -\bymin
    \divide \tpanelwidth by \tpdif
  \else \tpanelwidth=\panelwidth \fi
  \epsfig{file=#1,silent=,%
     bbllx=\bxmin bp,bblly=\bymin bp,bburx=\bxmax bp,bbury=\bymax bp,clip=,%
     width=\tpanelwidth mm,height=\tpanelheight mm}}
\def\labelypanel #1{
  \ifnum\panelheight=0 
    \tpdif=\bymax \advance\tpdif by -\bymin
    \multiply \tpdif by \panelwidth
    \tpanelheight=\tpdif
    \tpdif=\bxmax \advance\tpdif by -\bxmin
    \divide \tpanelheight by \tpdif
  \else \tpanelheight=\panelheight \fi
  \ifnum\panelwidth=0 
    \tpdif=\bxmax \advance\tpdif by -\bxmin
    \multiply \tpdif by \panelheight
    \tpanelwidth=\tpdif
    \tpdif=\bymax \advance\tpdif by -\bymin
    \divide \tpanelwidth by \tpdif
  \else \tpanelwidth=\panelwidth \fi
  \tpdif=\bxmax \advance\tpdif by -\tbxmin
  \multiply \tpanelwidth by \tpdif
  \tpdif=\bxmax \advance\tpdif by -\bxmin
  \divide \tpanelwidth by \tpdif
  \epsfig{file=#1,silent=,%
    bbllx=\tbxmin bp,bblly=\bymin bp,bburx=\bxmax bp,bbury=\bymax bp,%
    clip=,width=\tpanelwidth mm,height=\tpanelheight mm}}
\def\labelxpanel #1{%
  \ifnum\panelheight=0 
    \tpdif=\bymax \advance\tpdif by -\bymin
    \multiply \tpdif by \panelwidth
    \tpanelheight=\tpdif
    \tpdif=\bxmax \advance\tpdif by -\bxmin
    \divide \tpanelheight by \tpdif
  \else \tpanelheight=\panelheight \fi
  \ifnum\panelwidth=0 
    \tpdif=\bxmax \advance\tpdif by -\bxmin
    \multiply \tpdif by \panelheight
    \tpanelwidth=\tpdif
    \tpdif=\bymax \advance\tpdif by -\bymin
    \divide \tpanelwidth by \tpdif
  \else \tpanelwidth=\panelwidth \fi
  \tpdif=\bymax \advance\tpdif by -\tbymin
  \multiply \tpanelheight by \tpdif
  \tpdif=\bymax \advance\tpdif by -\bymin
  \divide \tpanelheight by \tpdif
  \epsfig{file=#1,silent=,%
    bbllx=\bxmin bp,bblly=\tbymin bp,bburx=\bxmax bp,bbury=\bymax bp,%
    clip=,width=\tpanelwidth mm,height=\tpanelheight mm}}
\def\labelxypanel #1{%
  \ifnum\panelheight=0 
    \tpdif=\bymax \advance\tpdif by -\bymin
    \multiply \tpdif by \panelwidth
    \tpanelheight=\tpdif
    \tpdif=\bxmax \advance\tpdif by -\bxmin
    \divide \tpanelheight by \tpdif
  \else \tpanelheight=\panelheight \fi
  \ifnum\panelwidth=0 
    \tpdif=\bxmax \advance\tpdif by -\bxmin
    \multiply \tpdif by \panelheight
    \tpanelwidth=\tpdif
    \tpdif=\bymax \advance\tpdif by -\bymin
    \divide \tpanelwidth by \tpdif
  \else \tpanelwidth=\panelwidth \fi
  \tpdif=\bxmax \advance\tpdif by -\tbxmin
  \multiply \tpanelwidth by \tpdif
  \tpdif=\bxmax \advance\tpdif by -\bxmin
  \divide \tpanelwidth by \tpdif 
  \tpdif=\bymax \advance\tpdif by -\tbymin 
  \multiply \tpanelheight by \tpdif
  \tpdif=\bymax \advance\tpdif by -\bymin
  \divide \tpanelheight by \tpdif
  \epsfig{file=#1,silent=,%
    bbllx=\tbxmin bp,bblly=\tbymin bp,bburx=\bxmax bp,bbury=\bymax bp,%
    clip=,width=\tpanelwidth mm,height=\tpanelheight mm}}

\def\CC{\par \vspace*{-2ex} \footnotesize \baselineskip=8pt \begin{verbatim}}

\long\def\startignore #1\stopignore{}   

\def\setlistparams{         
  \topsep=0.7ex                 
  \itemsep=0.7ex                
  \leftmargini=3ex}             
\setlistparams                  

\newcounter{alistindex}       

\newcounter{romenumnr}

\newlength{\minipagewidth}

\newsavebox{\boxcontent}
\newcommand{\ovalhead}[1]{
  \unitlength=1cm
  \sbox{\boxcontent}{\mbox{~~{#1}~~}}
  \begin{center}
    \ifdim\wd\boxcontent>6ex 
    \ifdim\wd\boxcontent<8cm 
    \begin{picture}(8,3) \thicklines     
      \put(4.0,0.8){\oval(8,1.6)} 
      \put(0.0,0.7){\parbox{8cm}{
         \begin{center} \usebox{\boxcontent} \end{center}}}
    \end{picture}
    \else \ifdim\wd\boxcontent<12cm 
    \begin{picture}(12,3) \thicklines     
        \put(6.0,0.8){\oval(12,1.6)} 
        \put(0.0,0.7){\parbox{12cm}{
           \begin{center} \usebox{\boxcontent} \end{center}}}
    \end{picture}
    \else
    \begin{picture}(16,3) \thicklines     
        \put(8.0,0.8){\oval(16,1.6)} 
        \put(0.0,0.7){\parbox{16cm}{
           \begin{center} \usebox{\boxcontent} \end{center}}}
    \end{picture}
    \fi \fi \fi
  \end{center}} 

\newcounter{headnr}            
\newcounter{subheadnr}[headnr]
\newcounter{subsubheadnr}[subheadnr]

\font\dropfont= cmr12 scaled \magstep5
\def\dropcap#1#2{{\noindent
    \setbox0\hbox{\dropfont #1}\setbox1\hbox{#2}\setbox2\hbox{(}%
    \count0=\ht0\advance\count0 by\dp0\count1\baselineskip
    \advance\count0 by-\ht1\advance\count0by\ht2
    \dimen1=.5ex\advance\count0by\dimen1\divide\count0 by\count1
    \advance\count0 by1\dimen0\wd0
    \advance\dimen0 by.25em\dimen1=\ht0\advance\dimen1 by-\ht1
    \global\hangindent\dimen0\global\hangafter-\count0
    \hskip-\dimen0\setbox0\hbox to\dimen0{\raise-\dimen1\box0\hss}%
    \dp0=0in\ht0=0in\box0}#2}

\def\rmit#1{{\it #1}}              
           
\def\ie{\rmit{i.e.,}}              
\def\eg{\rmit{e.g.,}}              

\def\specchar#1{\uppercase{#1}}    

\def\CaII{\mbox{Ca\,\specchar{ii}}}

\def\HI{\mbox{H\,\specchar{i}}} 
 
\def\MgII{\mbox{Mg\,\specchar{ii}}}

\def\Lyalpha{\mbox{Ly$\hspace{0.2ex}\alpha$}}
\def\Lybeta{\mbox{Ly$\hspace{0.2ex}\beta$}}

\def\CaIIH{\mbox{Ca\,\specchar{ii}\,\,H}}

\def\HK{\mbox{H\,\&\,K}}

\def\hk{\mbox{h\,\&\,k}}

\def\level #1 #2#3#4{$#1 \: ^{#2} \mbox{#3} ^{#4}$}   

\def\rme{{\rm e}}

\def\rmr{{\rm r}}

\def\={\hbox{$\!=\!$}}                     

\def\mathstacksym#1#2#3#4#5{\def#1{\mathrel{\hbox to 0pt{\lower 
    #5\hbox{#3}\hss} \raise #4\hbox{#2}}}}

\mathstacksym\lta{$<$}{$\sim$}{1.5pt}{3.5pt} 
\mathstacksym\gta{$>$}{$\sim$}{1.5pt}{3.5pt} 
\mathstacksym\lrarrow{$\leftarrow$}{$\rightarrow$}{2pt}{1pt} 
\mathstacksym\lessgreat{$>$}{$<$}{3pt}{3pt} 

\def\nl{,\ }
\def\edt#1{#1}
\def\psinu{\ensuremath{\psi (\nu,\vec{n})}} 
\def\phinu{\ensuremath{\phi (\nu,\vec{n})}} 
\def\be{\begin{equation}}
\def\ee{\end{equation}}
\def\ni{\ensuremath{n_i}} 
\def\nj{\ensuremath{n_j}} 
\def\dd{\mathrm{d}}
\def\dnud{\ensuremath{\Delta \nu_\mathrm{D}}}
\def\rii{\ensuremath{R_{\mathrm{II}}}}

\def\gii{\ensuremath{g_{\mathrm{II}}}}
\def\jg{\ensuremath{J_{\mathrm{r}}}}
\def\Oslo{Institute of Theoretical Astrophysics, University of Oslo\nl 
         P.O. Box 1029, Blindern\nl N--0315 Oslo\nl Norway}  
\def\UUt{Utrecht University\nl Postbus 80\,000\nl
         NL--3508 TA Utrecht\nl The Netherlands}
\def\ARC{NASA Ames Research Center\nl Mof{}fett Field\nl CA~94035\nl
  USA}
\def\LPARL{Lockheed Martin ATC\nl Solar \& Astrophysics Lab\nl 
         Org.\ H1--12, Bldg.\ 252\nl 3251 Hanover Street\nl 
         Palo Alto, CA~94304--1187\nl USA}
\def\SPO{NSO/Sacramento Peak\nl P.O. Box 62\nl 
         Sunspot, NM 88349--0062\nl USA}

\begin{document}

\title{Fast approximation of angle-dependent partial
  redistribution in moving atmospheres}

\titlerunning{Approximation of angle-dependent PRD in moving atmospheres}
  
\subtitle{}

   \author{J.~Leenaarts \inst{1,2}
   \and
   T. Pereira \inst{3,4}
   \and
   H. Uitenbroek \inst{5}
   }

   \offprints{ J. Leenaarts, \\ \email{jorritl@astro.uio.no} }

   \institute{\Oslo \and \UUt \and \ARC \and \LPARL \and \SPO}
   
   \date{Received; accepted}

   \abstract
       {}
       {Radiative transfer modeling of spectral lines including
       partial redistribution (PRD) effects requires the evaluation of
       the ratio of the emission to the absorption profile.  This
       quantity requires a large amount of computational work if one
       employs the angle-dependent redistribution function, which
       prohibits its use in 3D radiative transfer computations with
       model atmospheres containing velocity fields. We aim to provide
       a method to compute the emission to absorption profile ratio
       that requires less computational work but retains the effect of
       angle-dependent scattering in the resulting line profiles.}
       {We present a method to compute the profile
       ratio that employs the angle-averaged redistribution function
       and wavelength transforms to and from the rest frame of the
       scattering particles. We compare the emergent line profiles
       of the \MgII\,k and \Lyalpha\ lines computed with
       angle-dependent PRD, angle-averaged PRD and our new method in
       two representative test atmospheres.}
   {The new method yields a good approximation of
         true angle-dependent profile ratio and the resulting
         emergent line profiles while keeping the computational speed
         and simplicity of angle-averaged PRD theory.}  
{}

  
   \maketitle

\section{Introduction}                          \label{sec:intro}

The effects of partial frequency redistribution (PRD) are important
for spectral lines that form in environments with a low mass density,
where the average time between collisions between atoms and electrons
is long compared to the lifetime of an excited atomic state. Examples
are the \CaII\ \HK, \MgII\ \hk\ and Lyman lines of hydrogen in the
solar spectrum. Forward modeling of such lines requires a radiative
transfer algorithm that includes PRD effects.

In this paper we describe an improvement to the Multilevel
Accelerated Lambda Iteration (MALI) scheme of
\citet{1991A&A...245..171R,1992A&A...262..209R}
that has been extended to include PRD effects by
\citet{2001ApJ...557..389U}.
This improvement allows the scheme to treat PRD effects in atmospheres
with strong velocity fields using the angle-averaged redistribution
function without having to resort to the more accurate, but
computationally expensive, angle-dependent redistribution function. 

With this new method it is possible to accurately compute emergent
line profiles including PRD effects much faster than with the full
angle-dependent PRD treatment. This will allow detailed comparison of
radiation-MHD models with observations of chromospheric lines.

\section{Computation of partial redistribution effects}                          \label{sec:alg}

The solution of the radiative transfer problem with the MALI scheme
requires the evaluation of the ratio of the emission ($\psi$) and
absorption ($\phi$) profiles of a spectral line for which PRD effects
are important. Ignoring cross-redistribution, this ratio in the
observer's frame is given by
\citep[see, \eg][]{2001ApJ...557..389U}:

\begin{eqnarray}
 \label{angdepprd}
\rho(\nu,\vec{n}) &=& \frac{\psinu}{\phinu} \nonumber \\
 &=& 1 + \gamma \frac{\ni B_{ij}}{\nj P_j} \int
\oint \left( \frac{\rii(\nu,\vec{n},\nu',\vec{n}')}{\phi(\nu,\vec{n})}
- \phi(\nu',\vec{n}') \right) \nonumber \\ 
& & I(\nu',\vec{n}')
\, \frac{\dd \Omega'}{4 \pi} \, \dd \nu'. 
\end{eqnarray}
Here, $\nu$ and $\nu'$ are frequencies of the emitted and absorbed
photon in the observer's frame, $\vec{n}$ and $\vec{n}'$ their
respective directions, $\gamma$ is the coherency fraction, $n_i$ and
$n_j$ are the \edt{populations of the} lower and upper level of the line,
$B_{ij}$ is the
Einstein coefficient for radiative excitation and $P_j$ is the total
rate coefficient out of the upper level, $\rii$ is the angle-dependent
\edt{observer's frame} redistribution function for a broadened upper and
a sharp lower level and $I$ is the intensity.

The redistribution function is given by
\citet{1962MNRAS.125...21H}
as
\begin{eqnarray}
\rii(q,\vec{n},q',\vec{n}') &= & \frac{g(\vec{n},\vec{n'})}{4 \pi^2 \sin
  \gamma}
\exp \left[ - \left( \frac{q-q'}{2} \right) \csc^2 \left(\frac{\gamma}{2}\right)
\right] \\
& &  H \left( \frac{a}{\dnud} \sec \frac{\gamma}{2}, \frac{q+q'}{2} \sec \frac{\gamma}{2} \right). \nonumber
\end{eqnarray}
Here $g(\vec{n},\vec{n'})$ is the dipole scattering phase function,

\be
g(\vec{n},\vec{n'})=\frac{3}{16 \pi} \left( 1+\cos^2 \gamma \right),
\ee
$q^{(}{}'^{)}=(\nu^{(}{}'^{)}-\nu_0)/\dnud$ are the reduced frequencies of
the absorbed and emitted photon, with $\nu_0$ and $\dnud$ the line
center frequency and the Doppler width. The acute angle between
$\vec{n}$ and $\vec{n}'$ is given by $\gamma$, $H$ is the Voigt function and and $a$ the line
damping parameter.

Note that the second term in the integral of Eq.~\ref{angdepprd} is
\be
\int \oint \phi(\nu, \vec{n}) I(\nu,\vec{n}) \, \frac{\dd \Omega}{4
  \pi} \, \dd \nu
= R_{ij}/B_{ij},
\ee
with $R_{ij}$ the upward radiative rate coefficient that appears in the
rate equations.

In a static atmosphere the absorption profile does not depend on
direction. If one furthermore assumes that the radiation field is
isotropic, \ie
\be
 I(\nu,\vec{n}) = \frac{1}{4 \pi} \oint I(\nu, \vec{n}) \, \dd \Omega = J(\nu),
 \label{eq:isotropic}
\ee
then the angle integral in Eq.~\ref{angdepprd} can be performed
analytically
\citep[see][]{1982JQSRT..27..593H}
yielding a significantly simpler expression for the profile ratio:
\be \label{angindepprd}
\rho(\nu,\vec{n}) = 1 + \gamma \frac{\ni B_{ij}}{\nj P_j} \int
 \left( \gii(\nu,\,\nu') - \phi(\nu') \right) J(\nu')
 \, \dd \nu'.
\ee
This simplification is often called angle-averaged PRD.

The \edt{function} $\gii$ can be computed
efficiently using the approximation by \citet{1986A&A...160..195G}
with the generalization to cross-redistribution by
\citet{1989A&A...216..310U}. 

The assumption of an isotropic radiation
field generally yields results very close to the full angle-dependent
case for atmospheres without, or with only a weak, velocity field
\citep{2002ApJ...565.1312U}.

However, Eq.~\ref{angindepprd}\ typically becomes inaccurate when the
velocities in the atmosphere are larger than the Doppler velocity
$\sqrt{2kT/m}$ \edt{because Eq.~\ref{eq:isotropic} is no longer valid,}
and one should ideally use the angle-dependent formula
(Eq.~\ref{angdepprd}). Unfortunately, using the angle-dependent
expression is computationally expensive in both memory requirements
and computing time. 
The computation of Eq.~\ref{angdepprd} is at least a factor $
n_\mathrm{a}$ slower than the computation of
Eq.~\ref{angindepprd}, with $n_\mathrm{a}$ the number of rays used
for the direction integration of the radiation field. 

The way the \edt{angle-dependent} MALI scheme is set up one needs to
store the intensity
$I(\nu,\vec{n})$ and the profile ratio $\rho(\nu,\vec{n})$ in memory,
amounting to $2 \, n_\nu \, n_\mathrm{a}$ floating point numbers
per spatial grid point, in addition to all other quantities that need to be
retained. Ideally one also stores the redistribution function $\rii$
to avoid recomputation every iteration, which requires an additional
$n_\nu^2 \, n_\mathrm{a}^2$ floating point numbers per spatial grid point.

The memory and computing time requirements are easily met for a 1D
model atmosphere. However, if one wants to compute synthetic line
profiles from time series of 3D radiation-MHD models, a faster algorithm to
compute the angle-dependent PRD profile ratio is
desirable, if not required. 

\section{Hybrid PRD} \label{sec:hybrid}

\edt{The} algorithm \edt{we propose} should not be much slower than the
angle-averaged case,
yet capture the effects of velocities. It turns out that such a method
is fairly simple:

The agreement between angle-dependent and angle-averaged PRD in static
atmospheres implies that the angle-dependence caused by anisotropy of
the radiation field is minor. \edt{Instead, the majority of the
radiation anisotropy at a given wavelength experienced by
a parcel of gas in the observer's frame
in a moving atmosphere is caused by sampling the line profile at different
Doppler shifts when it receives radiation from different directions.
If we first transform the radiation field to the rest frame
of the moving gas parcel, we can again assume that the radiation field
is isotropic. Then} we can use Eq.~\ref{angindepprd} to
compute the profile ratio in the rest frame, \edt{replacing}
the angle-averaged intensity in the observer's frame
$J(\nu)$ with the angle-averaged intensity in the \edt{parcel's} rest
frame:
\be
\rho_\rmr(\nu_\rmr,\vec{n}) = 1 + \gamma \frac{\ni B_{ij}}{\nj P_j} \int
 \left( \gii(\nu_\rmr,\,\nu_\rmr') - \phi(\nu_\rmr') \right) J_\rmr(\nu_\rmr')
 \, \dd \nu_\rmr'.
\ee
\be \label{eq:Jgas}
\jg(\nu_\rmr)= \frac{1}{4 \pi} \oint  I(\nu_\rmr (1-\vec{n} \cdot \vec{u}/c), \vec{n}) \, \dd \Omega.
\ee
The quantity $\jg(\nu_\rmr)$ can be computed incrementally from
\edt{the observer's frame intensity}
$I(\nu, \vec{n})$ during the standard angle-frequency integration
needed to compute the radiative rate coefficients, without having to
keep the intensity for each spatial location, frequency and angle in
memory.

Plugging $\jg(\nu_\rmr)$ into Eq.~\ref{angindepprd} instead of
$J(\nu)$ yields a direction-independent emission to absorption ratio
in the rest frame $\rho_\rmr(\nu_\rmr)$.  Both $\jg(\nu_\rmr)$
and $\rho_\rmr(\nu_\rmr)$ must be stored in memory, but as they are
angle-independent they require a factor $n_\mathrm{a}$ less
storage than the angle-dependent quantities $I(\nu,\vec{n})$ and
$\rho(\nu,\vec{n})$. The MALI method requires $\rho$ in the observer's
frame, which is accomplished
through a frequency shift:
\be \label{eq:hybridrho}
\rho(\nu,\vec{n})=\rho_\rmr(\nu_\rmr (1+\vec{n} \cdot
\vec{u}/c)).  
\ee 
Numerically, this shift can be performed by a fast
interpolation.

\section{Results}                          \label{sec:results}

\begin{figure}
  \includegraphics[width=\columnwidth]{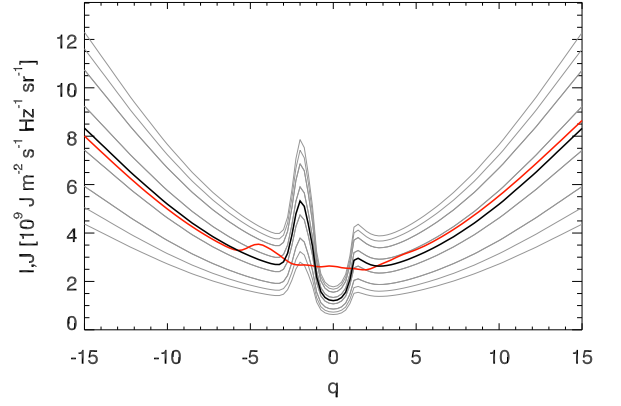}
  \caption{Plots of the radiation field used in the test
    computation as function of the reduce frequency $q$. Grey curves:
    intensity for different directions in the observer's frame.
     Black curve:
    angle-averaged radiation field in the observer's frame. Red curve:
    angle-averaged radiation field in the rest frame of a gas parcel
    moving with a velocity $(u_x,u_y,u_z)=(0,1,3)$ in units of
    the Doppler speed.
  \label{fig:jcomp}}
\end{figure}

\paragraph{Idealized test case}
To demonstrate the effect of the various methods to compute the
profile ratio we performed an idealized test case computation.

As intensity we took
the emergent intensity of the \CaIIH\ line computed from the standard
FALC model atmosphere
\citep{1990ApJ...355..700F}
with an artificially enhanced K2R peak to introduce an asymmetry in
the line profile. We also introduced
a directional anisotropy by setting
\be
I(\vec{n})= I_\rme (1-\frac{\cos \theta}{2}) 
\ee 
with $I_\rme$ the emergent intensity and $\theta$ the angle between
the vertical and the direction $\vec{n}$. These intensities are shown
as grey curves in Fig.~\ref{fig:jcomp} as function of the reduced
frequency $q= (\nu-\nu_0)/\dnud$ for different values of $\theta$,
where we assumed $\dnud = (\nu_0/c)\times8$\,km\,s$^{-1}$. The black
curve shows the angle-averaged intensity in the observer's frame
$J$. The red curve shows \jg, the angle-averaged intensity in the rest
frame of a gas parcel moving with a velocity of
$(u_x,u_y,u_z)=(0,1,3)$ in units of the Doppler speed. The angle
integrations were performed using the A8 set from
\citet{carlson1963},
which uses 10 rays per octant.
The presence of the velocity
smoothes the K2V and K2R peaks in the gas parcel's rest frame. The K2R
peak is reduced in height and has an increased width, the K3 minimum
is smoothed out and the K2V peak is a barely discernible hump at $q=4$.

Figure~\ref{fig:rhocomp} shows $\rho(q)$ given the intensity and gas
velocity described above for two different directions (given in the
upper left corner of the panels). We assumed $\gamma=0.9$ and $ \ni
B_{ij} / (\nj P_j) = 5 \times 10^5$. The black curve represents the
angle-dependent case (AD-PRD, Eq.~\ref{angdepprd}), the blue curve the
angle-averaged case (AA-PRD, Eq.~\ref{angindepprd}), while the red curve
shows the hybrid case (H-PRD, Eq.~\ref{eq:hybridrho}). The upper panel
displays scattering into the direction of movement and shows a
blue-shifted $\rho$-profile for the AD-PRD and H-PRD cases. The AA-PRD case
does not take the gas velocity into account and is centered at
$q=0$. In AD-PRD and H-PRD the profile ratio has a fairly smooth lowest
part, reflecting the smoothing of the radiation field in the rest
frame. Compared to H-PRD, the AD-PRD profile ratio exhibits additional
wiggles owing to the directional anisotropy of the radiation.

The lower panel shows scattering with an
angle close to perpendicular to the velocity. As a consequence all
$\rho$ profiles are approximately centered around $\rho=0$. Still the
H-PRD case reproduces the true AD-PRD profile ratio much better than the
AA-PRD formulation.

\begin{figure}
  \includegraphics[width=\columnwidth]{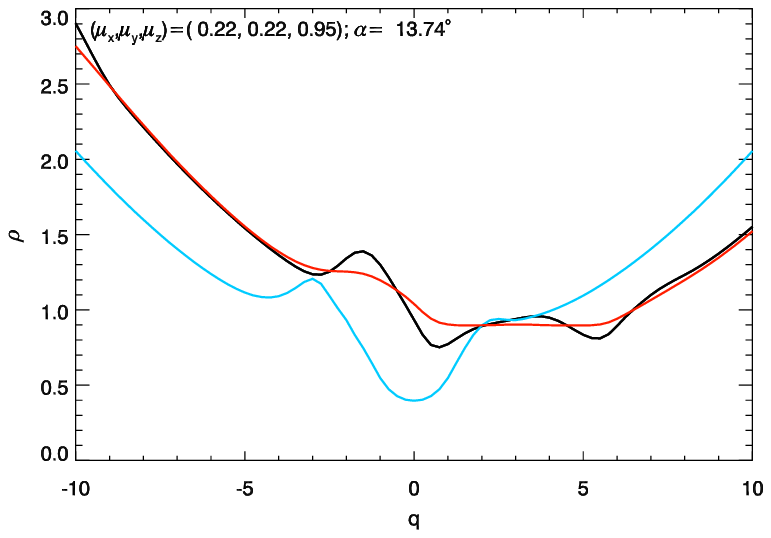}
  \includegraphics[width=\columnwidth]{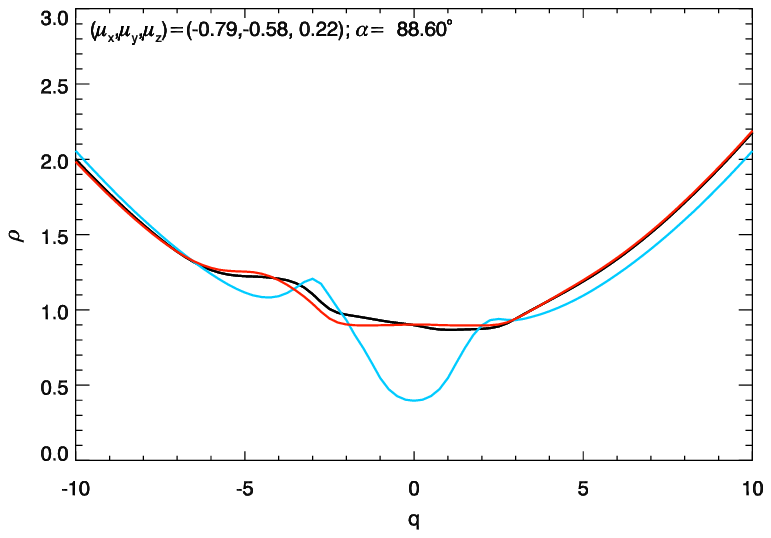}
  \caption{Emission to absorption profile ratio computed using
    angle-dependent redistribution (black, Eq.~\ref{angdepprd}),
    angle-independent redistribution (blue, Eq.~\ref{angindepprd})
    and our intermediate method (red, see Sec.~\ref{sec:alg}) in the
    observer's frame as function of reduced frequency $q$.  The cosines
    of the angle of the ray direction with the coordinate axes ($\mu_x,\mu_y,\mu_z$)  are
    given in the upper left corner of the panel. The quantity $\alpha$ gives the angle
    between the gas velocity and the ray direction.
  \label{fig:rhocomp}}
\end{figure}

\paragraph{1D test atmospheres}

We implemented the hybrid method into the RH code by
\citet{2001ApJ...557..389U},
which is also capable of using the angle-dependent and
angle-independent redistribution function. We used this code to
compute the emergent intensity  from five 1D plane
parallel model atmospheres in non-LTE with the three recipes for
PRD for a 4-level plus continuum \MgII\ and a 5-level plus continuum
\HI\ model atom.
As
model atmospheres we took columns from a snapshot of a 3D
radiation-MHD simulation computed with the Bifrost code
\citep{2011A&A...531A.154G}.
The details of this particular snapshot can be found in 
\citet{2012arXiv1202.1926L},
but are not important here: the snapshot merely provides
self-consistent model atmospheres with a velocity field. No microturbulence
was added. We treated \MgII\ \hk, \Lyalpha\ and
\Lybeta\ in PRD, all other lines were treated using complete
redistribution. For the angle integration of the radiation field we
used a ten-point Gauss-Legendre quadrature.

For all atmospheres and ray directions we found that the H-PRD method
reproduces the true AD-PRD emergent line profile much more accurately than the
AA-PRD method. We illustrate this in Fig.~\ref{fig:Icomp}, which shows the emergent intensity for a
near vertical ray ($\mu_z =0.95$) for the three different treatments
of the redistribution function for two different model atmospheres.

Both atmospheres show a complex velocity and temperature structure
(panels a and b). The second row (c and d) displays the emergent \MgII\ k
intensity. For this line we expect strong angle-dependent PRD effects
as the Doppler width of the line profile (2.6~km\,s$^{-1}$ at a temperature of
10~kK) is small compared to the flow velocities in the
atmospheres. This is indeed the case, the AA-PRD profile (blue) is very
different from the AD-PRD and H-PRD profiles. In panel c the former has a
higher emission peak at the blue side of the line core, whereas the
latter two have a higher peak on the red side. In panel d the AD-PRD and
H-PRD profiles show a stronger emission peak on the red side of the
line core.

The bottom row (e and f) compares the \Lyalpha\ profiles. The effect of
angle-dependent PRD is much smaller than for the \MgII\ k line because
the Doppler width of the line profile (13~km\,s$^{-1}$ at a temperature of
10~kK) is larger than the typical atmospheric velocities.

\begin{figure*}
  \includegraphics[width=\textwidth]{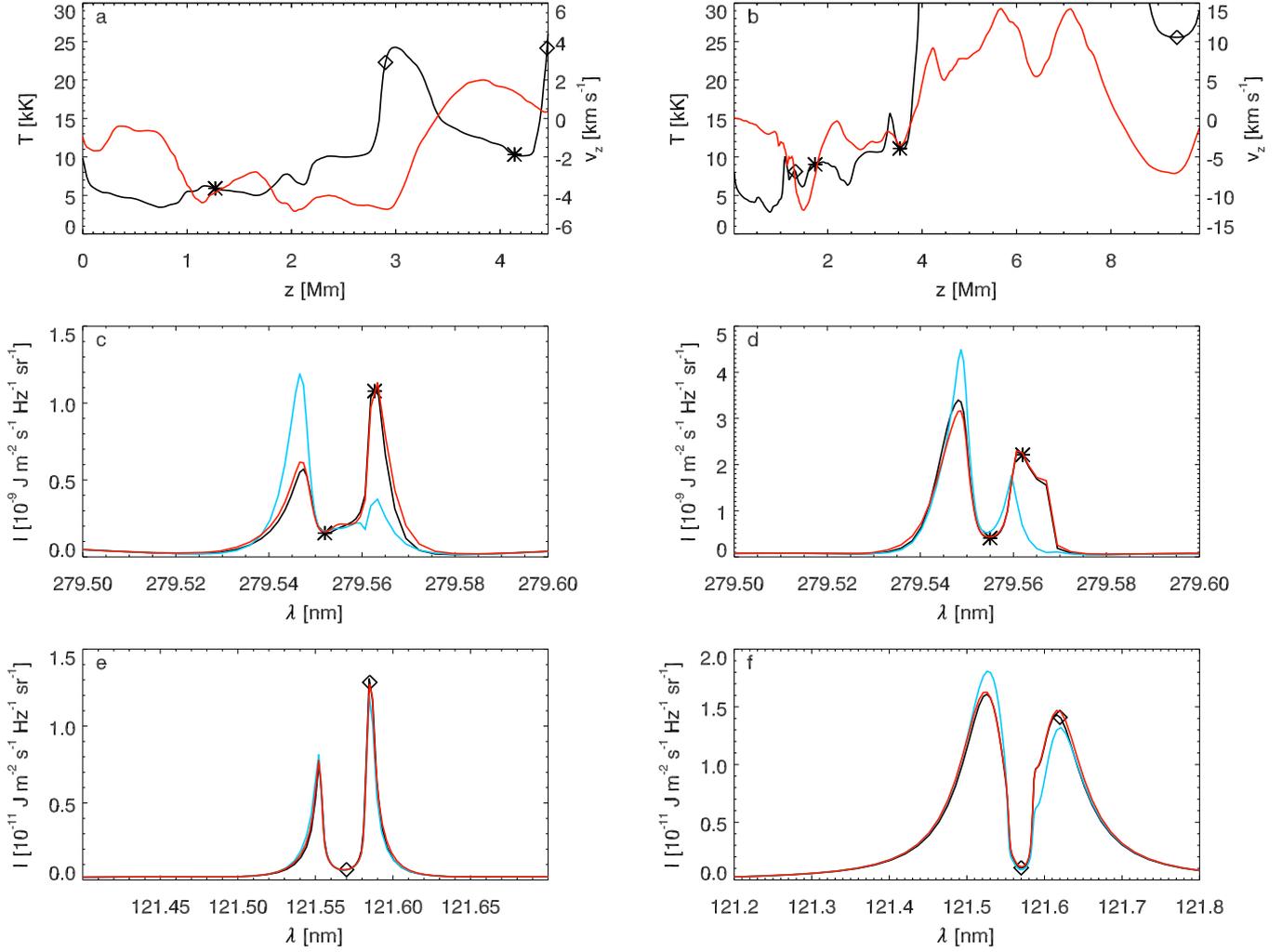}
  \caption{Comparison of PRD treatment in two plane-parallel model atmospheres
    (left-hand and right hand column respectively) taken
    from a radiation-MHD model. Panels a and b show the temperature
    (black) and vertical velocity structure (red) of the atmospheres. Panels c
    and d show the emergent intensity in the \MgII\, k line, with PRD
    treated fully angle-dependent (black), angle-averaged (blue) and
    the hybrid approach (red). Panels e and f show a similar profile
    comparison, but now for the \Lyalpha\ line of hydrogen. The stars
    and diamonds in panels a and b indicate the $\tau=1$ height for
    the wavelengths indicated in panels c, d, e and f, with stars
    for the  \MgII\, k line and diamonds for \Lyalpha.
  \label{fig:Icomp}}
\end{figure*}

\paragraph{Computational speed}

\begin{table}
\caption{Comparison of running time of PRD computations.}
\label{table:speed}
\centering
\begin{tabular}{rrrrrrr}
\hline\hline 
  & $n_\nu$ & $n_\nu^\mathrm{PRD}$ & & $t_\mathrm{FS}$ (s) &  $t_\mathrm{PR}$ (s) & $t_\mathrm{MALI}$  (s)\\
\hline 
\MgII & 354 & 198 & AA-PRD & 0.087 &  0.12 &   0.88  \\
      & 354 & 198 & AD-PRD & 0.096 & 40.00 & 121.00  \\
      & 354 & 198 & H-PRD  & 0.099 &  0.12 &   0.94  \\
\HI   & 491 & 148 & AA-PRD & 0.073 &  0.06 &   0.77  \\
      & 491 & 148 & AD-PRD & 0.067 &  9.40 &  29.00  \\
      & 491 & 148 & H-PRD  & 0.092 &  0.06 &   0.86  \\
\hline
\end{tabular} 
\end{table}

In Table~\ref{table:speed} we compare the running time of the code for
the three PRD methods for both model atoms in a 1D atmosphere with 225
spatial points (the atmosphere in panel a of
Fig.~\ref{fig:Icomp}). The quantities $n_\nu$ and $n_\nu^\mathrm{PRD}$
are the number of frequencies used, and the number of frequencies in
the PRD lines, respectively. The quantity $t_\mathrm{FS}$ is the time
to perform the formal solution for all frequencies and angles,
including the computation of $J_\mathrm{r}$ in the H-PRD case;
$t_\mathrm{PR}$ is the time to compute the profile ratios
(Eqs.~\ref{angdepprd},~\ref{angindepprd} and~\ref{eq:hybridrho});
$t_\mathrm{MALI}$ is the time to perform one full MALI iteration,
including three PRD sub-iterations
\citep[see][]{2001ApJ...557..389U}.

The time spent in the formal solution is slightly longer in the H-PRD
case compared to the other methods. This is due to the interpolations
required to compute Eqs.~\ref{eq:Jgas} and~\ref{eq:hybridrho} numerically.
The computing time of the AD-PRD method is mainly spent in the
computation of the profile ratio. The computation of the profile ratio
is several hundred times faster in the AA-PRD and H-PRD cases. 
The total time per MALI iteration for AD-PRD is a factor 130 (30) longer for the \MgII\
(\HI) computation than in the corresponding AA-PRD computation. In
contrast, the time per iteration for the H-PRD case is only $\approx$10\%
larger than for AA-PRD.

\section{Discussion \& conclusions} \label{sec:conclusions}

We have presented a fast approximate method to compute the
angle-dependent emission to absorption profile ratio needed to compute
line profiles for which PRD effects are important. Line profiles
computed with this method approximate the true profiles
computed with full angle-dependent PRD very well. 

Test computations show that the new hybrid method is 30 to 130 times
faster than the angle-dependent PRD computation and only $\approx$10\%
slower than the angle-averaged PRD method. This makes it possible to
compute accurate line profiles from time-series of 3D radiation-MHD models for lines
where angle-dependent PRD effects are important.

\begin{acknowledgements}
 JL recognizes support from the Netherlands Organization for
 Scientific Research (NWO).
\end{acknowledgements}

\bibliographystyle{aa} 
\bibliography{%
carlson,%
carlsson,%
fontenla,%
gouttebroze,%
hubeny,%
hummer,%
leenaarts,%
rybicki,%
uitenbroek%
}

\end{document}